\documentclass[graybox]{svmult}

\usepackage{pgfplots}

\usepackage{mathptmx}          
\usepackage[cmex10]{amsmath}

\usepackage{helvet}            
\usepackage{courier}           
\usepackage{type1cm}           

\usepackage{makeidx}           

\usepackage{graphicx}          
\graphicspath{{./figNum/}}

\usepackage{epstopdf}
\usepackage{caption}
\usepackage{subcaption}

\usepackage{multicol}          
\usepackage[bottom]{footmisc}  

\usepackage{pdfpages}

\usepackage{listings}

\usepackage{hyperref}
\usepackage{longtable}
\usepackage{siunitx}
\usepackage{csquotes}
\usepackage{framed}
\usepackage{comment}
\usepackage{lscape}
\usepackage{rotating}
\usepackage[font=small,labelfont=bf,tableposition=top]{caption}
\usepackage{booktabs,multirow}

\usepackage{tikz}
\usetikzlibrary{shapes,arrows}
\usepackage{xcolor}
\usepackage{colortbl}

\definecolor{mygreen}{RGB}{28,172,0} 
\definecolor{mylilas}{RGB}{170,55,241}
\definecolor{timberwolf}{rgb}{0.86, 0.84, 0.82}
\makeindex             

\begin{document}

\title*{A Comparison of Turbulence Generated by 3DS Sparse Grids With Different Blockage Ratios 
and Different Co-Frame Arrangements}

\titlerunning{3DS Spare Grid Turbulence}

\author{M. Syed Usama$^{a}$ and Nadeem A. Malik$^{a,b*}$ }
\authorrunning{Usama \& Malik}

\institute{$^{*}$ Corresponding author e-mail: nadeem.malik@ttu.edu, nadeem\_malik@cantab.net
\and
\at $^a$  King Fahd University of Petroleum and Minerals,  Dhahran, Saudi Arabia \\
$^b$ Department of Mechanical Engineering, Texas Tech University, Lubbock, TX 74909, USA }

\maketitle

\abstract{ {
A new type of grid turbulence generator, the 3D sparse grid (3DS),  is a co-planar arrangement of co-frames 
each containing a different length scale of grid elements [Malik, N. A.  US Patent No. US 9,599,269 B2 (2017)]
and possessing a much bigger parameter space than the flat 2D fractal square grid (2DF). Using DNS we compare the 
characteristics of the turbulence (mean flow, turbulence intensity, energy spectrum) generated by different 
types of 3DS grids.  The peak intensities generated by 3DS can exceed  the peaks generated by the 2DF by 
80\%;  we observe that a 3DS with blockage ratio 24\% produces turbulence similar to the 2DF with blockage 
ratio 32\% implying lower energy input for the same turbulence.}}

\section{Introduction}\label{intro}
\label{sec:1}

\begin{figure}[t]   
\begin{center}
\begin{subfigure}[t]{0.3\linewidth}
\centering
  \includegraphics[width=4.5cm]{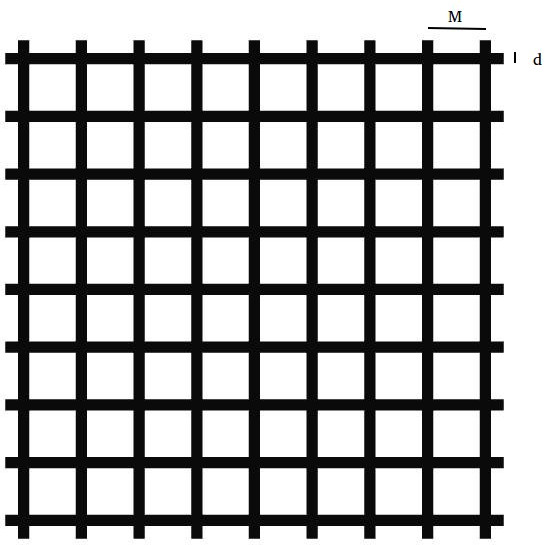}
  \caption{Regular grid (RG)}        \label{fig1a}
\end{subfigure} \hspace{17mm}
\begin{subfigure}[t]{0.36\linewidth}
  \centering
  \includegraphics[width=5cm]{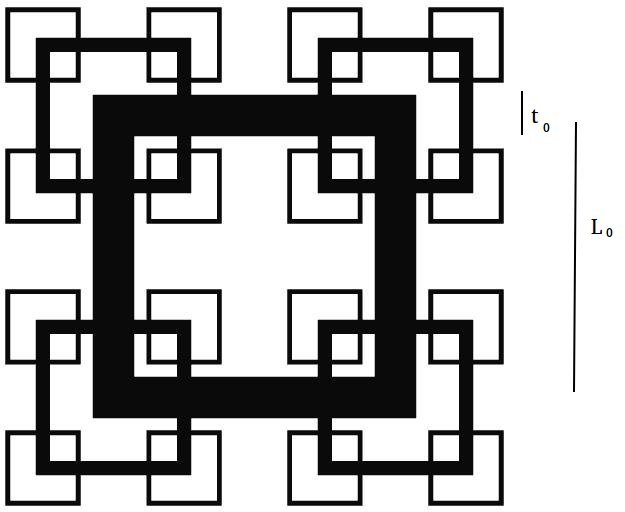}
  \caption {Square fractal grid (2DF)}   \label{fig1b}
\end{subfigure}\\
\vspace{10mm}
\hspace{-40mm}
\begin{subfigure}[t]{0.3\linewidth}
  \centering
  \includegraphics[width=8cm]{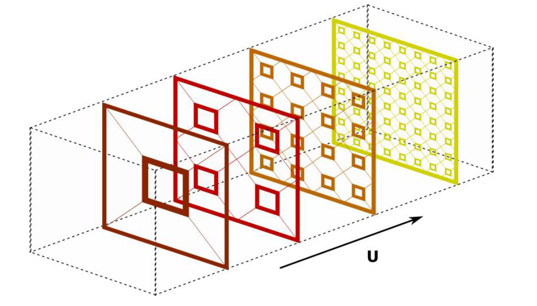}
  \caption {3D sparse grid (3DS)}   \label{fig1c}
\end{subfigure}
  \caption {Different types of grid}
\end{center}
\end{figure}

{The generation and control of turbulence is one of the most important challenges in fluid mechanics, with applications ranging from drag reduction to mixing in chemical reactors.
A promising innovation in recent times has been the design of a new types of turbulence generating grids which are different to the classical regular grid (RG), Fig. 1(a). The RG grids have bars of fixed thickness and flow passages of fixed size. A new grid type is a multi-scale arrangement of bars of varying thicknesses that produce flow passages of various sizes. Typically, the bar thicknesses and flow passages are in a self-similar configuration in a two-dimensional plane, such as the 2D square fractal grid (2DF), Fig. 1(b). A key feature of multi-scale 2D grids is that they produce multiple scales of turbulence at once in the grid plane, which alters the turbulence generated \cite{Queiros2001,Seoud2007,Laizet2011,Vassilicos2015,Sakai2016} compared to RG; in particular the peak turbulence intensity is enhanced for the same blockage ratio \cite{Laizet2011}.}

Conisder a rectangular flow channel or conduit with a turbulence grid placed close to the entrance. A defining charactersitic of turbulence generating grids is the blockage ratio (i.e. the solidity), $\sigma$, which is the surface area of all the bar elements, $S_{elts}$ divided by the planar cross-sectional area $A$ of the channel,
\begin{eqnarray}
   \sigma &=& \frac{S_{elts}}{A}
\end{eqnarray}
In the RG and 2DF, $\sigma$ is a single value; in \cite{Laizet2011} a three-generation 2DF was presented with $\sigma_{2DF}=0.32$ (or 32\%). {$\sigma$, the solidity, is important for flow passage; for the same volumetric flow rate you need higher pressure gradient in a channel with higher solidity, which means more energy input requirement. Thus, an important goal in mixing is to optimize the balance between energy input (or $\partial P/\partial x$), the solidity $\sigma$, and turbulence generation.}

A recent innovation in grid generated turbulence, the Sparse 3D Multi-Scale Grid Turbulence Generator, or 3D sparse grid (3DS) for short \cite{Malik1,Malik2}, has excited interest in the turbulence community because of its potential to alter and control  turbulence characteristics even more than the 2DF. The 3DS separates each generation of length scale of grid elements into its own co-frame in overall co-planar arrangement, Fig. 1(c), which produces a 3D ‘sparse’ grid system. Each generation of grid elements produces a turbulent wake pattern that interacts with the other wake patterns downstream. The length scale of the grid elements from co-frame to co-frame can be in any geometric ratio, although a fractal pattern across the generations is a popular choice. 
{The spacing between successive co-frames $r_1, r_2, ...$ are new parameters which do not exist in a non-sparse single frame 2D grid system. If each co-frame if located at $[x_0,x_1,x_2, ...]$, then $r_1=x_1-x_0$, $r_2=x_2-x_1$, etc. Each co-frame has a blockage ratio, $\sigma_{0}, \sigma_{1}, \sigma_{2}, ...$. 
We define the overall (or maximum) blockage ratio of the 3DS system,  $\sigma_{3DS}$, to be the maximum in this set of values, 
\begin{eqnarray}
  \sigma_{3DS} &=& Max\{\sigma_0, \sigma_1, \sigma_2, ...\}. 
\end{eqnarray}
Thus, for the same value of $\sigma_{3DS}$ there are an infinite number of possible 3DS configurations since the $\sigma_i's$ can take continuous values, provided $0<\sigma_i\le \sigma_{3DS}$; at least one (possibly all) of the co-frames must have $\sigma_i=\sigma_{3DS}$.}

A third new parameter in the 3DS grid system is the order of arrangement of the co-frames $Z_0, Z_1, Z_2, ...$ which can be in any order. We define $Z_0$ to be the largest scale of elements, $Z_1$ the next largest, and so on. Thus, a 3-generation 3DS grid system $[X_0,X_1,X_2]= [Z_0,Z_1,Z_2]$ where the co-frames are placed at, $[x_0,x_1,x_2]$ such that $x_0<x_1<x_2$, means that the co-frame length scales  $[l_0,l_1,l_2]$, are such that $l_0>l_1>l_2$. However, the 3DS grid system $[X_0,X_1,X_2]=[Z_1,Z_2,Z_0]$ where the co-frames are placed at, $[x_0,x_1,x_2]$ such that $x_0<x_1<x_2$, means that the co-frame length scales  $[l_0,l_1,l_2]$, are such that $l_2>l_0>l_1$.

It is imporant to note that $\sigma_{3DS}$ is much smaller than in the comparative 2DF grid, $\sigma_{3DS}\ll \sigma_{2DF}$. In a 3-generation 3DS system the blockage ratios of the three co-frames is $[\sigma_1, \sigma_2,\sigma_3]$, and with a geometric ratio of $a=0.5$ between the successive generation, we obtain  $\sigma_1 =\sigma_2 =\sigma_3 =\sigma_{2DF}/3$. Therefore, if the blockage ratio of the 2DF is $\sigma_{2DF}=30\%$, then  $\sigma_{3DS}=10\%$.  (This will differ for other   $a\not=0.5$ \cite{Usama2018}.)

{ We use Direct Numerical Simulations to compare the mean flows, the turbulence intensities, and the energy spectra generated by three-generation 3DS grid systems with different blocking ratios and different order of co-frame arrangements, and we also compare them to the turbulence produced by RG and 2DF grids. In this study we keep $r_1$ and $r_2$ constant. Here, our systems are channels with periodic lateral boundary conditions; the possible effects of no-slip wall conditions and of changing mean flow direction is discussed in Section 4, Conclusions.}


\section{Direct Numerical Simulations}
\label{sec:2}
In the first instance we compare the 3DS with the simuations of Laizet et al. \cite{Laizet2011}. The simulated domain has dimensions of $460.8\times115.2\times115.2 d_{min}^3$  where $d_{min}$  is the thickness of the smallest square. The height and width of the channel is  $H=115.2d_{min}$. 

The effective mesh size in the RG is  $M_{ef}=13.33d_{min}$ , and the bars have length $115.2d_{min}$, and thickness $2.6d_{min}$. This matches the system reported in  \cite{Laizet2011}. 

The 2DF has non-dimensionlized lengths and widths $\{l_i,d_i\}$ , in generation  $i=0,1,2$.  Where $l_0=57.6=0.5h$, $l_1=0.5l_0$, $l_2=0.5l_1$. The  bar thicknesses are  $d_0=8.5$,  $d_1=2.9d_2$,  $d_2=1$. All lengths are henceforth non-dimensionalized by  $d_{min}$.
The time scale is defined by  $t_2=d_{min}/U_\infty$ where $U_\infty$ is the inlet velocity set equal to $1$. 

The 3DS-2, Table 1, has the same lengths and thickness as the 2DF above, however each generation is held in a co-frame separated from the next by non-dimensional distances,   $r_1=x_1-x_0=17$, and  $r_2=x_2-x_1=8.5$, and  $x_0=10$, where $x_i's$  are the  non-dimensionalised $x$-coordinates of the i’th frame. 

The blockage ratio (or solidity)  in the RG and 2DF is the same 32\%. The maximum blockage ratio in the 3DS is $15\%$. 

\begin{table*}
\setlength{\tabcolsep}{12pt}
\renewcommand{\arraystretch}{1.5}
\begin{center}
\def~{\hphantom{0}}
{
}
\begin{tabular}{lll lll ll}
\hline \\
Grid	  &$X_0$   &$\sigma_0$ &$X_1$   &$\sigma_1$   &$X_2$   &$\sigma_2$ 	& $\sigma / \sigma_{3DS}$	\\
\hline \\
RG	 &-    &-	&-   &-   &-        &-       &32\%      \\[-2pt]
2DF	 &-    &-	&-   &-   &-        &-       &32\%       \\[-2pt]
3DS-2	 &$Z_0$  &15\%      &$Z_1$  &15\%    &$Z_2$  &15\%  &15\%  \\[-2pt]
3DS-3	 &$Z_0$  &24\%      &$Z_1$  &15\%    &$Z_2$  &15\%  &24\%  \\[-2pt]
3DS-4	 &$Z_0$  &32\%      &$Z_1$  &15\%    &$Z_2$  &15\%  &32\%  \\[-2pt]
3DS-5	 &$Z_1$  &15\%      &$Z_2$  &15\%    &$Z_0$  &32\%  &32\%  \\[-2pt]
3DS-6	 &$Z_2$  &15\%      &$Z_0$  &32\%    &$Z_1$  &15\%  &32\%  \\[-2pt]
&&& &&& &\\
\hline \\
\end{tabular}
 \caption{Different grid types used in this study: the order of arrangement of the co-frames $Z_i$ and the corresponding 
 co-frame blockage ratio $\sigma_i$ (\%) are shown. The last column shows the maximum (i.e. overall) blockage ratio 
 of the grid system. \protect}\label{Table1}
 \end{center}
 \end{table*}

OpenFOAM, (OFoam), was used to create a numerical grid $N_x\times N_y\times N_z = 2304\times 576\times 576$. The RG and 2DF  grids lie in the plane $x_0=10$  downstream of the channel inlet. Periodic boundary conditions were applied on the walls in the $y$  and $z$  directions; and inlet-outlet boundary conditions were applied in the  $x$-direction. The initial condition is a uniform inflow velocity  $U_\infty=1$. The Reynolds number is,  $Re=\frac{U_\infty d_{min}}{\nu}=300$. The resolution is  $\Delta x=0.2d_{min}$ which is adequate for our purposes. 

OpenFoam is 2nd order accurate in spatial resolution which is adequate for low Reynolds numbers. It uses finite volume discretization  with Pressure Implicit Splitting of Operator Algorithm (PISO). Time discretization using Backward Euler method, whereas gradient and Laplacian term discretization using Gauss linear method are performed. Divergence term discretization is done using Gauss cubic method which is a third order scheme. Interpolation and other terms are discretized using Gauss Linear schemes. The resulting linear systems are solved by preconditioned conjugate gradient method with diagonal incomplete Cholesky preconditioner for pressure solution whereas iterative solver is used with symmetric Gauss-Siedel as the smoother to calculate velocities. Tolerance is set at  $10^{-6}$. Simulation time step is  $\Delta t=0.015d_{min}/U_\infty$  which corresponds to a Courant number of  $0.75$. { Blockage, such as a bluff body, is achieved by imposing no-slip $u=0$ condition on the numerical grid corresponding to the surface of the body. The square cross-sectional bars in the 3DS are particularly easy to implement as they match exactly the rectangular geometry of the finite volume elements.}

\begin{figure}[!p]
\begin{center}
\hspace{-20mm}
\begin{subfigure}{0.3\textwidth}
  \includegraphics[width=5cm]{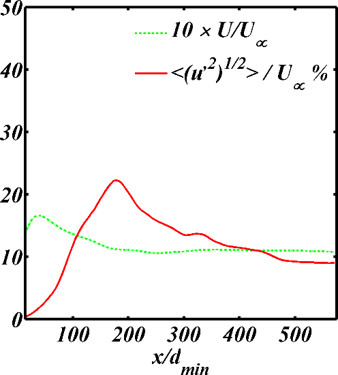}   
  \caption{2DF: $\sigma_{2DF}=32\%$} \label{fig2a}
  \end{subfigure}
\hspace{20mm}
\begin{subfigure}{0.3\textwidth}
  \includegraphics[width=5cm]{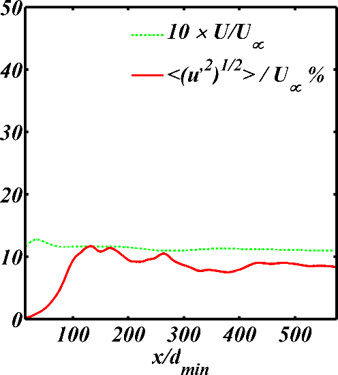}  
  \caption{3DS-2: $[Z_0,Z_1,Z_2]$; $\sigma_{Max}=15\%$} \label{fig2b}
\end{subfigure}\\
\vspace{5mm}
\hspace{-20mm}
\begin{subfigure}{0.3\textwidth}
  \includegraphics[width=5cm]{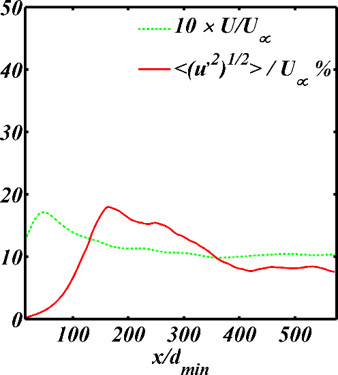}   
  \caption{3DS-3:   $[Z_0,Z_1,Z_2]$; $\sigma_{Max}=24\%$} \label{fig2c}
\end{subfigure}
\hspace{20mm}
\begin{subfigure}{0.3\textwidth}
   \includegraphics[width=5cm]{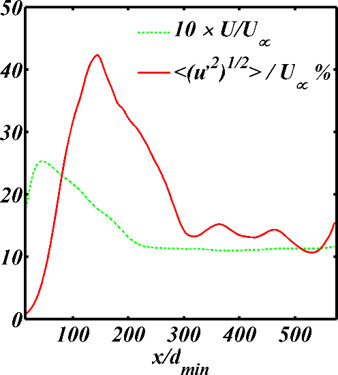}  
  \caption{3DS-4:  $[Z_0,Z_1,Z_2]$; $\sigma_{Max}=32\%$} \label{fig2d}
\end{subfigure}\\
\vspace{5mm}
\hspace{-20mm}
\begin{subfigure}{0.3\textwidth}
  \includegraphics[width=5cm]{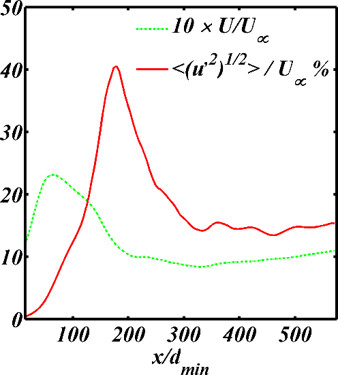}  
  \caption{3DS-5:  $[Z_1,Z_2,Z_0]$;  $\sigma_{Max}=32\%$} \label{fig2e}
\end{subfigure}
\hspace{20mm}
\begin{subfigure}{0.3\textwidth}
  \includegraphics[width=5cm]{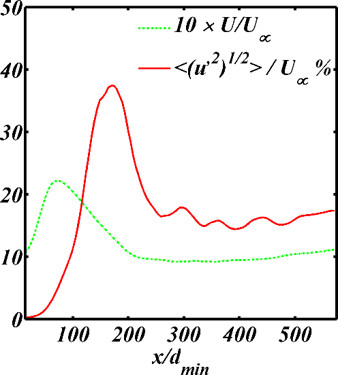}
  \caption{3DS-6:   $[Z_2,Z_0,Z_1]$;  $\sigma_{Max}=32\%$} \label{fig2f}
\end{subfigure}\\
\caption{The mean streamwise velocity $U/U_\infty$ (green), and the streamwise turbulence intensity $u'/U_\infty$ (red) along the centerline, from the 2DF and the 3DS grids. The 3DS co-frame order of arrangement $[Z_i]$, and the blockage ratios $\sigma_{2DF}$ and $\sigma_{3DF}$ (\%) are shown.}
\end{center}
\end{figure}

\begin{figure}[!p]
\begin{center}
\hspace{-25mm}
\begin{subfigure}{0.3\textwidth}
  \includegraphics[width=5.8cm]{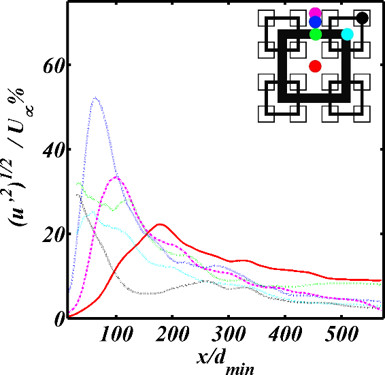}   
  \caption{2DF: $\sigma_{2DF}=32\%$} \label{fig3a}
\end{subfigure}
\hspace{23mm}
\begin{subfigure}{0.3\textwidth}
  \includegraphics[width=5.8cm]{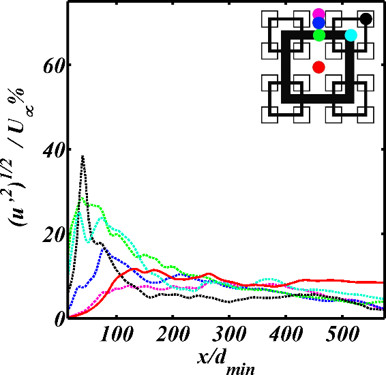}   
  \caption{3DS-2: $[Z_0,Z_1,Z_2]$; $\sigma=15\%$} \label{fig3b}
  \end{subfigure}\\
\vspace{5mm}
\hspace{-25mm}
\begin{subfigure}{0.3\textwidth}
  \includegraphics[width=5.8cm]{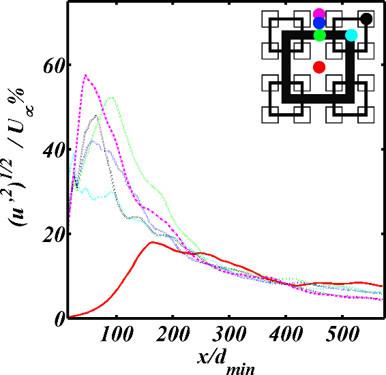}   
  \caption{3DS-3: $[Z_0,Z_1,Z_2]$; $\sigma=24\%$} \label{fig3c}
\end{subfigure}
\hspace{23mm}
\begin{subfigure}{0.3\textwidth}
  \includegraphics[width=5.8cm]{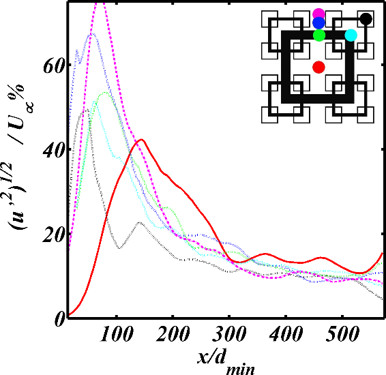} 
  \caption{3DS-4: $[Z_0,Z_1,Z_2]$; $\sigma=32\%$} \label{fig3d}
\end{subfigure}\\
\vspace{5mm}
\hspace{-25mm}
\begin{subfigure}{0.3\textwidth}
  \includegraphics[width=5.8cm]{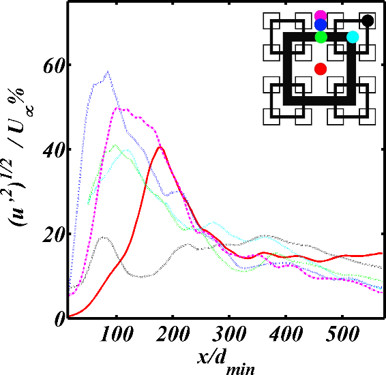}  
  \caption{3DS-5: $[Z_1,Z_2,Z_0]$; $\sigma=32\%$} \label{fig3e}
\end{subfigure}
\hspace{23mm}
\begin{subfigure}{0.3\textwidth}
  \includegraphics[width=5.8cm]{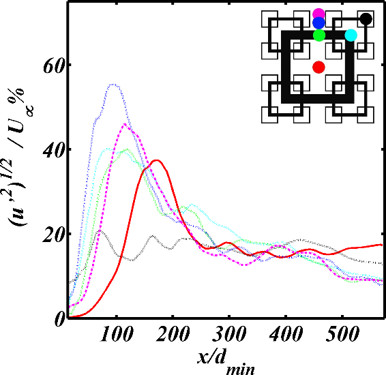}   
  \caption{3DS-6: $[Z_2,Z_0,Z_1]$; $\sigma=32\%$} \label{fig3f}
\end{subfigure}\\
\caption{The streamwise turbulence intensity$u'/U_\infty$ along different pencils as indicted, from the 2DF and the 3DS grids. The 3DS co-frame order of arrangement $[Z_i]$, and the blockage ratios $\sigma_{2DF}$ and $\sigma_{3DF}$ (\%) are shown.}
\end{center}
\end{figure}

\begin{figure}[!p]
\begin{center}
\hspace{-25mm}
\begin{subfigure}{0.3\textwidth}
  \includegraphics[width=5.8cm] {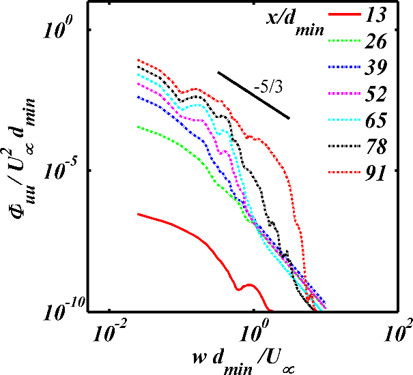} \label{fig4a}
  \caption{2DF: $\sigma=32\%$} 
\end{subfigure}
\hspace{23mm}
\begin{subfigure}{0.3\textwidth}
  \includegraphics[width=5.8cm]{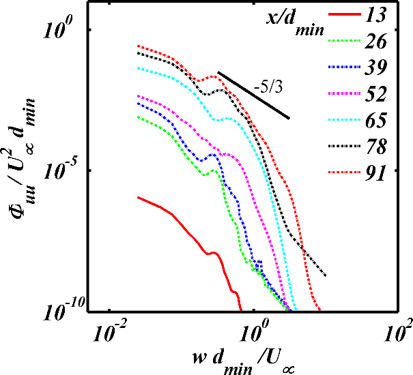}
  \caption{3DS-2: $[Z_0,Z_1,Z_2]$; $\sigma=15\%$} \label{fig4b}
\end{subfigure}\\
\vspace{5mm}
\hspace{-25mm}
\begin{subfigure}{0.3\textwidth}
  \includegraphics[width=5.8cm]{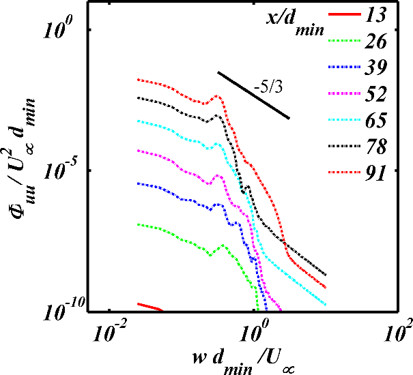}  
  \caption{3DS-3: $[Z_0,Z_1,Z_2]$; $\sigma=24\%$} \label{fig4c}
\end{subfigure}
\hspace{23mm}
\begin{subfigure}{0.3\textwidth}
  \includegraphics[width=5.8cm]{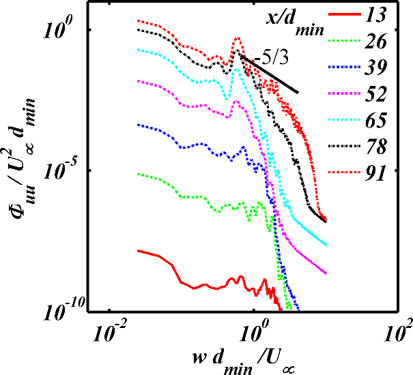}  
  \caption{3DS-4: $[Z_0,Z_1,Z_2]$; $\sigma=32\%$} \label{fig4d}
\end{subfigure}\\
\vspace{5mm}
\hspace{-25mm}
\begin{subfigure}{0.3\textwidth}
  \includegraphics[width=5.8cm]{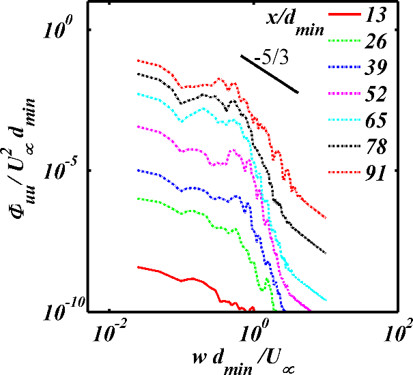}
  \caption{3DS-5: $[Z_1,Z_2,Z_0]$; $\sigma=32\%$} \label{fig4e}
\end{subfigure}
\hspace{23mm}
\begin{subfigure}{0.3\textwidth}
  \includegraphics[width=5.8cm]{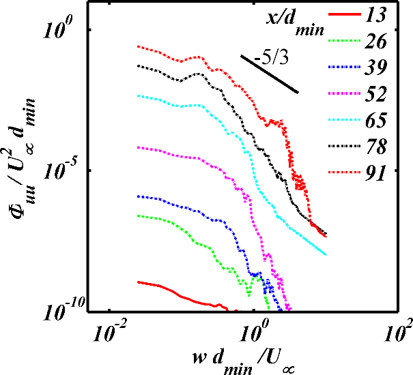}
  \caption{3DS-6: $[Z_2,Z_0,Z_1]$; $\sigma=132\%$} \label{fig4f}
\end{subfigure}\\
\caption{The energy spectrum $\Phi_{uu}/U^2_\infty d_{min}$ against the wavenumber $k=wd_{min}/U_\infty$, at different locations along the centerline, from the 2DF and the 3DS grids. The 3DS co-frame order of arrangement $[Z_i]$, and the blockage ratios  
$\sigma_{2DF}$ and $\sigma_{3DF}$ (\%) are shown.}
\end{center}
\end{figure}


\section{Results on turbulence intensity}
\label{sec:3}

A comparison of the turbulence intensities along different pencils from the RG, 2DF, and 3DS-2 grids from DNS simulations has been reported in \cite{Usama2018}. The RG and 2DF plots are close to the results in \cite{Laizet2011} which validates the DNS for these calculations. 

Here, in Fig. 2 we show the time averaged mean flow along the centerline, $U(x=0)/U_\infty$, from all six grids considered in Table 1, and the centerline time averaged rms turbulence  fluctuation (i.e. intensity), $u'(x=0)/U_\infty$. 

Fig. 3 shows the turbulence intensity, $u'(x)/U_\infty$, from the same grids along different pencils  in the x-direction as indicated.

We group the results into three sets for comparison in  Fig. 2 and Fig. 3: the first set is (a) and (b), where the 2DF and 3DS-2 are compared. 3DS-2 is obtained from the 2DF by taking the grid bars in 2DF and placing them in the different  co-frames. 

The second set is (b), (c), and (d), which is a comparison of 3DS grids with different blockage ratios for the same co-frame arrangement, $[Z_0,Z_1,Z_2]$. 

The third set is (c), (d), and (f), which is a comparison of 3DS grids with different co-frame arrangements for the same blockage ratio $\sigma_{3DS}=32\%$. 

As expected, for the low blockage ratio 3DS-2 $\sigma_{3DS}=15\% \ll \sigma_{2DF}$ the mean flow along the centreline in the 3DS-2 is not much disturbed, and the turbulence intensity generated remains low at $\approx 10\%$. However, away from the centreline, the turbulence intensity shows significant peaks in the near field close to the grid, although not as much as in the 2DF. In all cases, in the far field downstream the planar averaged turbulence intensity decays slowly. Thus, it is in the near to mid-range downstream where the differences are most strongly felt.

The results in Figs. 2(c) and 3(c), from the 3DS-3 grid with $\sigma_{3DS}=24\%$ are remarkably close to the 2DF (32\%), Figs. 2(a) and 3(a). The mean and the intensity along the centerline are similar, and the off-centerline turbulence intensities in Fig. 3(c)  display similar trends as well, the peak intensities being only a little higher along most of the pencils. 

Figs. 2(d) and 3(d), from the 3DS-4 with $\sigma_{3DS}=32\%$, show the peaks in mean flow and the turbulence intensities exceeding the 2DF peaks by as much as 80\% in the near-field downsteam. The peaks in Figs. 3(d) are the highest yet observed.

The comparison of the order of arrangement of the co-frames in the 3DS grids for $\sigma_{3DS}=32\%$, (d)--(f), shows that the turbulence is sensitive to the ordering, although not as sensitive as a change in $\sigma_{3DS}$. The three cases are a cyclic permutation, with the largest scale $Z_0$ being cycled. The order $[Z_0,Z_1,Z_2]$ in the 3DS-4, Fig. 3(d), shows the highest peaks in turbulence intensity, although the other two cyclic cases 3DS-5 and 3DS-6 also produce higher peaks that the 2DF. The peaks in the mean flow do not differ much, all three cases being about 50\% higher than in the 2DF.

We note that in some of the 3DS grid cases the centerline  mean flow and turbulence intensity appear to {\em increase} far downstream towards the end of the channel. This is almost certainly due to the entrainment of turbulent flows towards the center of the channel, because the current 2DF and 3DS are void of elements in the center. (Other geometric configurations may produce different results.)

Finally,  Fig. 4 shows the energy spectrum at different locations downstream for all the grids considered. The spectra are obtained from time series of the velocities at the given location, and converted from frequency domain to the wavenumber domain, $\Phi_{uu}(k)$, where $k\sim wd_{min}/U_\infty$, using Taylor's hypothesis. The 2DF approaches equilibrium turbulence, $\Phi_{uu}\sim k^{-5/3}$ the fastest, and most of the 3DS cases do not achieve this till around $x/d_{min}\approx 100$  remaining in non-equilibrium because the turbulence is still developing in this region. The 3DS appears to prevent a return to equilibrium more effectively than other types of grid.


\section{Conclusions}\label{Discussion}
The three-generation 3DS grids that we have investigated show remarkable sensitivity to the blockage ratio $\sigma_{3DS}$ and the order of arrangement of co-frames when compare to the 2DF grid. Our results show that the three-generation 3DS-3 grid with $\sigma_{3DS}=24\%$ with co-frame ordering $[Z_0,Z_1,Z_2]$ produces turbulence characteristics that are close to the 2DF with $\sigma_{2DF}=32\%$; if this could be translated to lower pressure gradient (i.e. lower energy input) then this would be very significant for industrial applications. Furthermore, the senstivity of the turbulence to the grid parameters implies that a better way of controling the turbulence generated could be devised.
The 3DS grids with blockage ratio equal to the 2DF -- 3DS-4, 3DS-5, and 3DS-6, with $\sigma_{3DS}=\sigma_{2DF}=32\%$ in cyclic co-frame ordering respectively -- show peaks in the mean flow and the turbulence intensity in the near field downstream of the grid that greatly exceed that from 2DF grid, by as much as 80\% along some pencils.  
The turbulence spectra show that the turbuence generatd by the 3DS grids remain far from equilibrium for the longest period downstream. The entrainment of the turbulence toward the center of the channel causes the mean flow and the intensity to increase far downstream along the centerline.

{The results presented here constitute a proof of concept for the 3DS. As this is the first study in 3DS we have simplified the system to facilitate a direct comparison with the RG and 2DF of \cite{Laizet2011}; we have ignored the boundary wall effects which generates turbulence of its own that would penetrate towards the centre as the streamwise distance increases. However, if the 3DS grid is placed close to the channel entrance, then the effect of boundary walls may not be so important close to the centerline in the near field.  
It is also of some interest to speculate about how effective the 3DS would be in a bigger system where the mean velocity is changing directions. Shear generated turbulence will likely increase but would need greater pressure drop. On the other hand, if the mixing and turbulence characteristics are dependent mainly on the generation of length scales and time-delay between the co-frames, then it may not matter so much.  This and the effect of other parameters, such as varying the inter-frame distances, $r_1$ and $r_2$, is left for future investigation.
}

\section*{Acknowledgements}
The authors acknowledge the support  from King Abdullah University of Science and Technology (KAUST) for making available the High Performance Computing facility Shaheen 2 for this project.

\end{document}